# The Product Promotion and Consumer Retention Gap in Online Shopping

Senthur Balan S
Design Engineer
QuEST
Bangalore, India
senthurbalan@gmail.com

Sowmyan Jegatheesan
Managing Head
CyberWurl
Chennai, India
jsowmyan@gmail.com

Sakthi Ganesh M
Assistant Professor
VIT University
Vellore, India
Sakthiganesh.m@vit.ac.in

*Abstract—* As the number of online shopping websites increases day by day, so are the online advertisement strategies and promotional techniques. The number of people who uses internet keeps on increasing daily and it has become a vast marketplace to promote products, surely it will be a prime reason to drive any companies' growth in the future.. This paper primarily focuses on the areas on which online shopping lags product promotion and customer retention. Sellers must concentrate on the areas in which online marketing lags product promotion techniques; also they should introduce new strategies to increase their market share to gain customer's attention towards their products.

## I. INTRODUCTION

Ever since the internet has been made public, people have witnessed its effect in their daily life in one way or the other, with the commerce industry not being an exemption. From the opening of an online pizza shop by Pizza Hut to the online shopping site that was opened by Amazon back in 1995, the online shopping industry has grown exponentially. In the last decade the internet has witnessed a massive progress notably in the developing countries region and continues to attract more number of people from these regions towards it. Online marketing articulated with the second generation websites (i.e., blogs, forums, social networks, etc.,) has given the advantage to people for communicating about the various products and services they use in their day to day life (from mobile phones to the SUVs they use). Nowadays people don't necessarily have the time or energy to walk up and down the aisles of a 200,000 square foot super store to find out that one time [14]. Interpersonal communication is often an influential source of product information for consumers [8]. Even companies have also started using the internet as a platform to connect with their customers, to take surveys and promote their products. This shows us that social embeddedness has become the next step in product promotion for companies [13].

The present day market is completely driven by consumer choices as there are plenty of products to choose from; also the range of customers for a product keeps widening, according to American Demographics, "A woman is no longer simply a woman for marketing purposes. She is, for example, a single mom, an ethnic minority, a bicycling enthusiast, a buyer of petite-sized clothing, and a wine connoisseur" [14]. As a coin has got two sides, online shopping has also got its own advantage and disadvantage in promoting the products. Consumers must be well aware of the products they buy online in order to get the right product they are looking for, similarly people must also get to know the advantages of commercial marketing over online shopping and try to implement those techniques to promote online sales. This will result in long term buyer-seller relationship and improved returns for the company.

## II. THE HIGH POTENTIAL MARKETS FOR ONLINE SHOPPING

Like any other business the online market has also got the high potential and emerging markets, which could bring in huge margins for the e-commerce industry in the years to come. As we infer from the table of data, it is quite clear that the online shopping market will be mainly affected by the countries mentioned in the table shown below as the most number of internet users are from these countries only. As China and India are the top two most populous countries, they will be the high potential market; the internet penetration rate in those two countries is very low when compared to other countries in the list, so the potential of these countries' market remain untapped.

Moreover the e-commerce industry should be able to identify the exact demands of the online shopping community people and start to customize the websites and product promotion strategies according to the culture in these regions, since the culture in the Asian region will be entirely different from that of the western world. The key drivers of online shopping growth are mainly internet penetration, income levels and cultural factors.

TABLE 1. Internet Usage Data

| Rank | Country | Internet users | Percentage by population |
|---|---|---|---|
| 1 | China | 456,238,464 | 34.30 |
| 2 | US | 243,542,822 | 79.00 |
| 3 | India | 121,567,256 | 10.50 |
| 4 | Japan | 102,063,316 | 80.00 |
| 5 | Brazil | 81,748,504 | 40.65 |
| 6 | Germany | 66,825,986 | 81.85 |
| 7 | Russia | 59,937,788 | 43.00 |
| 8 | UK | 52,996,180 | 85.00 |
| 9 | France | 51,879,480 | 80.10 |
| 10 | Nigeria | 45,944,229 | 28.43 |

The per-capita income of the people in these two countries (India and China) is also increasing in a notable scale. Huge growth in online shopping is expected, led by china and India but with other countries such as Thailand also likely to exhibit strong growth-as incomes and internet penetration rise [10]. Another survey brings out the fact that 47% of the internet users of the country are avid shoppers, even though the country has the little internet penetration rate and very few percentage of the people buying online compared to its total population, the numbers are far ahead than rich nations like Australia, Hong Kong and Singapore [10].

### III. GOODS THAT ARE BOUGHT ONLINE MOSTLY

In the global level online shopping market, people mostly prefer buying books followed by clothing/accessories/shoes, airline ticket/reservations and electronic equipments [1]. The online shopping is mainly utilized by the younger generation, so the products that are sold online are also mostly focused towards this band of people. In order to bring in the elder people inside online shopping community, more products/services targeted towards them should also be made available online.

### IV. AREAS IN WHICH ONLINE SHOPPING LAGS PROMOTION

As compared with the commercial shopping, online shopping has few setbacks and those are listed here. Every factor has been explained in detail in the coming sections.
- Commercial stores vs. online stores
- Virtual marketing effects
- Improper product reviews
- Grievance responding mechanism

*B. Commercial stores vs. online stores*

A commercial store which has been designed to attract customers (either individual stores or the stores in shopping malls) will have all the aesthetic and functional values which will try to retain their customers in their stores for more time, this could attract the attention of the buyer towards the product and there are more chances that a buyer could buy it [15]. Due to the crowding and excitement available at the malls and in shops people will have more tendency of buying a product [7]. This is not the case in online shopping as individuals will be surfing the products from a remote machine and they will not get the real-time shopping feel in them. An individual who is shopping in a commercial store will have the chance to have a feel of the product's features in real, rather than reading the product's specs as in the case of online shopping. A person who has checked a product with physical senses has more chances of buying them than the one who is simply seeing them [6]. So a buyer going through a website will be having only less chances of buying a product rather than a person who is checking out that product in stores. As a person visits the store in real time he/she has the opportunity to get the technical advice regarding the usage of the product from the store people, which in case of online shopping people have to rely on other's comments and reviews on that product. In recent times some online shopping websites have started incorporating online help agents in the form of chat windows, via which the customer can get in touch with the technical support person to get details, but not all the online shopping/ service providers are offering this type of facility.

Since shopping in a commercial store will turn out to be a physical activity rather than seeming to be like browsing a page on the internet, chances are very high for the buyer to recollect that shopping experience and associate himself with the product and store. This would help the store or the brand of the product to enable the buyer to buy the same brand or buy from the same store. The allocation of shelf space is of vital importance to retailers, because it influences both inventory return on investment and customer satisfaction [3]. Another paper reveals that placing a product in the center of the shelf helps products gets noticed and, ultimately, bought [16]. These kind of influences are not possible in case of online shopping. Also the buyer-seller relationship has also got its share in enabling the buyer to buy the product; buyers who have a previous relationship with a seller may care about the seller's actions, whereas other buyers with no purchase relationship may not care about the seller's actions. Even when prior relationships exist, they could vary in terms of commitment, intimacy, satisfaction, and self-connection [2].

*B. Virtual marketing effects*

The theoretical and research contributions of the past two decades have established emotions as a legitimate area of scientific inquiry in the field of marketing [4]. Research shows that when people imagine either positive or negative things they used to overestimate things rather than what the real effects would be [2]. People frequently overestimate the impact of an event when they imagine it, relative to when they actually experience it. This phenomenon, known as the impact bias, has been well established as an interpersonal phenomenon [2]. As buyers glance through online shopping sites, they start to imagine things about how they will use the product and the benefits the product could offer. When multiple mobile models are compared through online sites they (websites) will list all the technical details of those mobiles in a tabular format. Now, if we take up the display screen dimensions for all the models, people will start to imagine the size of the display screen for each mobile. Even if there is only a minor change in the dimensions of the display screen for all the mobiles, buyers can't exactly predict it as how it would be in real. This will not be the case in commercial shopping as everyone can have a trial on the product they may buy. When simply imagining incidents and making predictions, people do not fully realize the extent to which their sense-making mechanisms will lower the emotional impact of actually occurring incidents, thus leading to an impact bias such that predictions overestimate the impact of an actual experience [2].

A study reveals the fact that people when asked to recommend the website from which they purchased products recently, people who were new to that website found more connected towards the website rather than the one who has used the website for many times. Also the new customer to the website feels more loyal to the website/service when he/she recommends the website to their friends or family members [5]. From this we can observe the fact that word of mouth is more effective in bringing in more customers, but the factor that rises concern is the people who are new to the website has more affinity towards the website rather than the one who shops from the website for a very long time. Only a long term customer will know much about the website, so again it is made clear that due to people's initial attraction towards a website, they could rate the website much higher and this could make other buyers think that website to be a good one, where in reality that website could be a average rated website too, this can also be classified under impact bias.

Due to the proliferation of advertisements on internet, people are exposed to wide range of ads, and most of the time the ad never reaches the targeted audience; this makes people to ignore ads and makes advertisers to lose money. Reaching the targeted audience in the case of online advertising is increasingly becoming complex as the details being provided by most of the people in websites is not accurate. This virtual marketing also makes the buyer to buy things what they like on the screen rather than what buyers really need for them. People who can't wait for a long time before having the product on their hand will not be opting for online shopping, as the delivery might take time. Privacy issues are becoming increasingly important in this information age. Another research finding points out that almost 15% of all Americans and 19% of all Internet users have been victims of credit card fraud. Furthermore, regardless of whether buyers have experienced such privacy violation incidents firsthand or not, the frequent coverage by popular media makes such incidents easy to imagine. This is an important factor that needs to be addressed to gain the goodwill of a buyer in order to establish a long-term buyer-seller relationship through online shopping sites.

*C. Improper product reviews*

An individual's decision on buying a product is highly influenced by the advice/suggestion he/she receives from their peers. A study has revealed that opinions are most important when it comes to purchasing consumer electronic items: 57% of online respondents consider reviews prior to buying. Reviews on cars (45%) and software (37%) rounded out the top three most important online influences when making a purchase [1]. Every one's perception towards a product will be different and so their experiences with the product would be. If we simply browse the reviews of a particular product on internet, we may end up with a heap of reviews and we might find it hard to gain the correct perspective about a product. In this we can't judge what a person was exactly expecting from a product and compare it with our needs from that product. Also in order to promote a product many fake and paid up reviews are appearing in the internet just to gain the attention of the viewer and to gain his good will for the product. Due to the vast nature of the internet many people are finding it hard to channelize their search and obtain the reviews of the products which are not much popular in the internet. The type of reviews will also change according to the geographical data of the person. An African user of a particular model of mobile phone will have a different opinion than an Asian user, since the features in a product will not definitely meet all the demands made by the different sector of people. Sometimes even people are getting confused to take decisions by reading the reviews as some are really biased about the nature of the product.

The internet posses to be a ground where we can check out the details of any product (microphone to sports cars), but it is unsure that we will get appropriate insight about the product according to our likes. Online comments and reviews about products will have no sense if a person is buying a product which is getting released for the first time. Online businesses lose as many as 67% of consumers due to lack of online product information [11].

*D. Grievance responding mechanism*

The next major source for competitive advantage likely will come from more outward orientation toward customers, as indicated by the many calls for organizations to compete on superior customer value delivery [12]. More than the quality of any product, the people are very much concerned about the services they would receive after buying that product. This is one of the reasons which are responsible to drive the sales figures of a Computer manufacturer; their excellent after market services even forced the other hardware manufacturing (IT related) companies to introduce similar concepts to retain their market share. Technical products have become increasingly complex, driving consumers to the internet for information before a purchase, importantly; consumers also rely on the internet for help in using technical products post-purchase [8]. So companies should make sure that they are present on the internet to monitor their buyer's demands, their comments and they need to address all those incidents within the stipulated time period else they will start losing the customer base. A finding discloses the fact that people in the Asia Pacific region consider that "Secure payment facility" is the most prioritized factor (87%) that could affect propensity of online shopping. Security possess to be the topmost concern even for the people in China (87%), Thailand (75%), Hong Kong (74%) and India (73%) [10]. So the online shopping websites should be ready to provide the utmost transaction and online security services in order to boost the customer's belief and retain them for long periods.

Research indicates that people who are overly irritated with a product or service only will be taking time to lodge a complaint rather than a one who simply dislikes the product [14]. So if a person buying a product through online is not getting back to the website or the company it doesn't mean that the user is satisfied with the product. People who are buying goods through online shopping may not get a chance to interact with the person who is actually selling it. Since the product will be delivered to the customer only through some courier service the user of the product will have no other chance other than lodging a complaint online and wait till the customer care people call him back. Not all online shopping service has a very good customer's grievance responding mechanism. So it is very clear that whichever company is providing the best after market service in the coming days will definitely be a game changer and wide range customer base generator.

## V. CONCLUSION

As the previous topics have indicated, the future of the product sales or product promotion will be mainly dependent up on the internet and its users. As companies are trying all the means to capture the people's attraction in order to promote their growth, they must be well aware of the areas in which online marketing lags its pace. So this paper has given the areas on which the concentration of the companies should be for promoting their growth. Further researches could be done by implementing new product promotion strategies on these areas and their results could be studied.